\documentclass{nature}

\usepackage{graphicx} 
\usepackage[margin=0.7in]{geometry}
\usepackage{amsmath}
\usepackage{hyperref} 
\usepackage{multirow}
\usepackage{newtxtext}         \usepackage{chemformula} 
\usepackage{lineno}
\usepackage{caption}
\usepackage{float}
\usepackage{booktabs}

\hypersetup{colorlinks, 
	linkcolor={blue!75!black!80!yellow},
	citecolor={blue!75!black!80!yellow}, 
	urlcolor={blue!75!black!80!yellow}
	}
\usepackage[cmintegrals]{newtxmath}
\usepackage{lipsum}
\usepackage{lineno}
\usepackage{soul,xcolor}
\setstcolor{red}
\newcommand{\BostonCollege}{Department of Physics, Boston College, Chestnut Hill, MA, USA}

\title{Uncovering the Hidden Ferroaxial Density Wave as the Origin of the Axial Higgs Mode in \ch{RTe3}}
\author{Birender Singh$^{1*}$,  Grant McNamara$^{1}$, Kyung-Mo Kim$^{1}$, Saif Siddique$^{2}$, Stephen D. Funni$^{2}$, Weizhe Zhang$^{3}$, Xiangpeng Luo$^{3}$, Piyush Sakrikar$^{1}$, Eric M. Kenney$^{1}$, Ratnadwip Singha$^{4}$, Sergey Alekseev$^{5}$, Sayed Ali Akbar Ghorashi$^{5}$, Thomas J. Hicken$^{6}$, Christopher Baines$^{6}$, Hubertus Luetkens$^{6}$, Yiping Wang$^{1}$, Vincent M. Plisson$^{1}$, Michael Geiwitz$^{1}$, Connor A. Occhialini$^{7}$, Riccardo Comin$^{7}$, Michael J. Graf$^{1}$, Liuyan Zhao$^{3}$, Jennifer Cano$^{5,8}$, Rafael M. Fernandes$^{9}$, Judy J. Cha$^{2}$, Leslie M. Schoop$^{4}$ and Kenneth S. Burch$^{1}$}

\begin{document}

\maketitle
\begin{affiliations}
    \item \BostonCollege
    \item Department of Materials Science and Engineering, Cornell University, Ithaca, NY, USA
    \item Department of Physics, University of Michigan, Ann Arbor, MI, USA
    \item Department of Chemistry, Princeton University, Princeton, NJ, USA
    \item Department of Physics and Astronomy, Stony Brook University, Stony Brook, NY, USA
    \item Laboratory for Muon Spin Spectroscopy, Paul Scherrer Institute, CH-5232 Villigen, Switzerland
    \item Department of Physics, Massachusetts Institute of Technology, Cambridge, MA, USA
    \item Center for Computational Quantum Physics, Flatiron Institute, NY, USA
    \item School of Physics and Astronomy, University of Minnesota, Minneapolis, MN, USA
   \end{affiliations}

\begin{abstract}
The recent discovery of an axial amplitude (Higgs) mode in the long-studied charge density wave (CDW) systems \ch{GdTe3} and \ch{LaTe3} suggests a heretofore unidentified hidden order. A theoretical study proposed that the axial Higgs results from a hidden ferroaxial component of the CDW, which could arise from non-trivial orbital texture. Here, we report extensive experimental studies on \ch{ErTe3} and \ch{HoTe3} that possess a high-temperature CDW similar to other \ch{RTe3} (R = rare earth), along with an additional low-temperature CDW with an orthogonal ordering vector. Combining Raman spectroscopy with large-angle convergent beam electron diffraction (LACBED), rotational anisotropy second-harmonic generation (RA-SHG), and muon-spin relaxation ($\mu$SR), we provide unambiguous evidence that the high-temperature CDW breaks translation, rotation, and all vertical and diagonal mirror symmetries, but not time-reversal or inversion. In contrast, the low-temperature CDW only additionally breaks translation symmetry. Simultaneously, Raman scattering shows the high-temperature CDW produces an axial Higgs mode while the low-temperature mode is scalar. The weak monoclinic structural distortion and clear axial response in Raman and SHG are consistent with a ferroaxial phase in \ch{RTe3} driven by coupled orbital and charge orders. Thus, our study provides a new standard for uncovering unconventional orders and confirms the power of Higgs modes to reveal them. 
\end{abstract}

In most cases, the symmetries broken by a centrosymmetric CDW state with a single-$q$ are entirely encoded in the wave vector $\vec{q}_{CDW}$. The magnitude of $\vec{q}_{CDW}$ gives the inverse period, reflecting the translational symmetry breaking. The orientation of $\vec{q}_{CDW}$ with respect to the atomic wavevectors may break point symmetries, such as rotational and mirror symmetries. Recently, attention has turned to unconventional CDW that break additional symmetries other than those encoded by $\vec{q}_{CDW}$ due to the simultaneous condensation of secondary ferroic order parameters\cite{Fradkin2015, Comin:2015aa, Tam2022, Liu2023, Luo2021, xu2020, Hart2023, Xi2015, yumigeta2021, PRB81, kwon2024, Hildebrand2018}. Identifying and understanding such systems is challenging, requiring uncovering which material properties are reflected in the order parameter. Beyond probe limitations, certain broken symmetries are often hidden due to the absence of a clear conjugate field to which they couple.

\begin{figure}[ht]
\centering\includegraphics[width=1\textwidth]{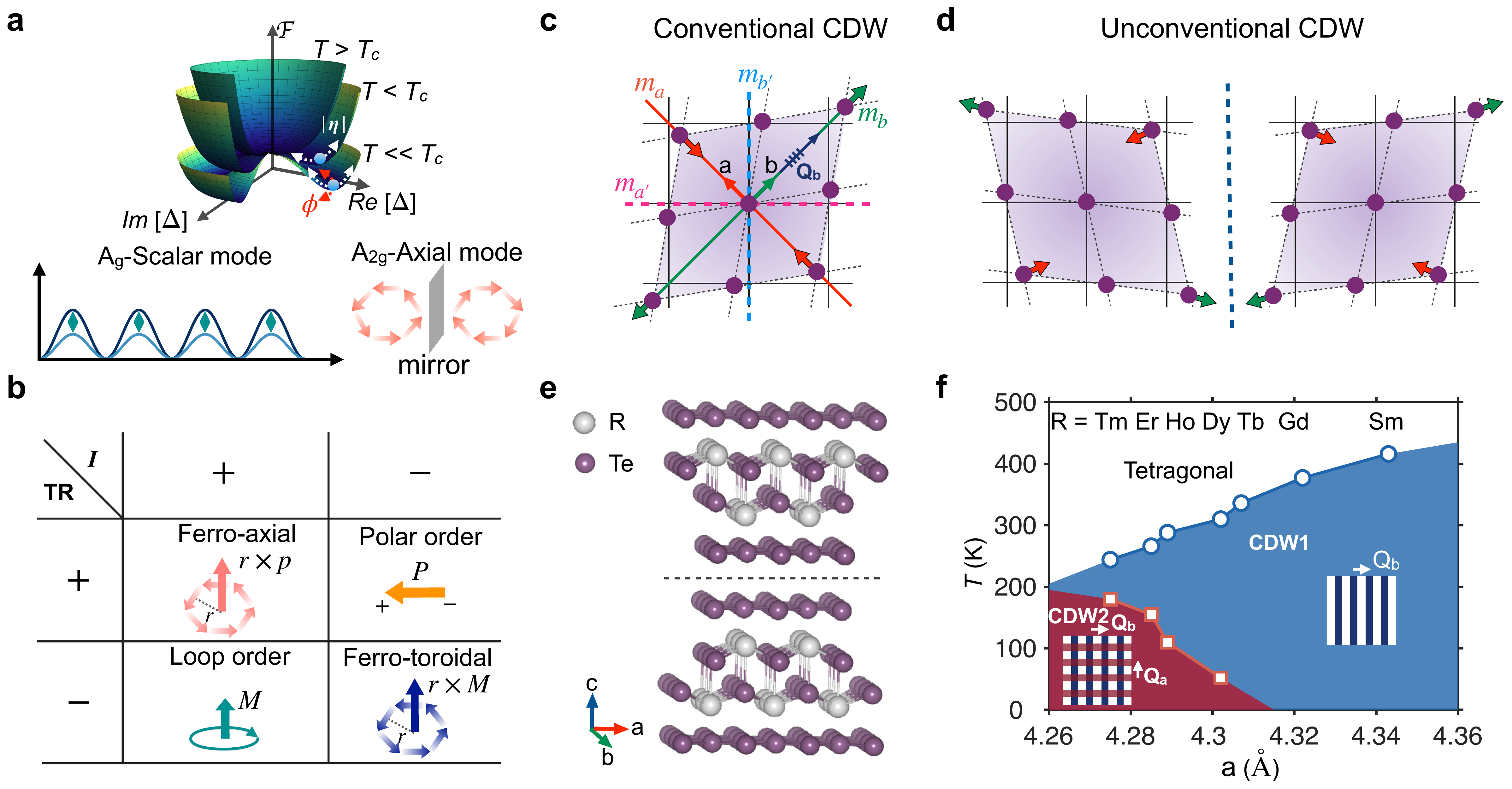}
\vspace{-0.31cm}
\caption{\textbf{Spontaneous symmetry breaking.} \textbf{a} Free energy versus CDW order parameter components (top), and a diagram of the conventional ($A_g$-scalar) and unconventional ($A_{2g}$-axial) CDW amplitude modes (bottom). \textbf{b} The ferroic order parameters possible under inversion (\textit{I}) and time-reversal (TR)\cite{van2007,Hayami2018}. \textbf{c} Schematic for the conventional unidirectional CDW distortion, breaking translation, \textit{C$_4$}-rotation, and diagonal mirrors (\textit{m$_{a'}$} (pink) and \textit{m$_{b'}$} (blue)) symmetries. Solid line represents unbroken vertical mirrors (\textit{m$_{a}$} (red) and \textit{m$_{b}$} (green)) plane, and dark blue arrow along $b$-axis represents CDW wave-vector (Q$_{b}$). Purple circle indicates the Te-atoms. \textbf{d} Unconventional (ferroaxial) CDW order breaking translation, rotation, and all vertical (\textit{m$_a$}, \textit{m$_b$}) and diagonal (\textit{m$_{a'}$}, \textit{m$_{b'}$}) mirror symmetries. \textbf{e} Crystal structure of \ch{RTe3} (dashed line indicates the glide plane). \textbf{f} Phase diagram of \ch{RTe3} as a function of lattice parameter. Inset shows the unidirectional (dark blue) and bidirectional (red) CDW states.}
   \label{fig:symmetry-breaking}
\end{figure}

These unconventional CDW orders could be observed and understood through their novel quasi-particles with properties inherited from the orders in which they emerge\cite{Gruner1994,Matsunaga2013,Xie2024}. Indeed, a new phase's fundamental broken symmetries (or topologies) should be reflected in their collective excitations' properties\cite{souliou2017,schwarz2020,puviani2021,wulferding2023}. Consider, for example, a conventional CDW described by a Ginzburg-Landau approach with an order parameter ($\psi=\eta e^{i\phi}$), where $\eta$ reflects the distortion amplitude and $\phi$ the phase with respect to the lattice. The resulting free energy of the ordered state (Fig.~\ref{fig:symmetry-breaking}a) produces two new low-energy excitations of the order parameter field (amplitudon (Higgs), and phason (Goldstone))\cite{Gruner1994}. Since $\psi \neq 0$ typically breaks symmetries determined by $\vec{q}_{CDW}$, the amplitude mode ($|\psi|^2$) transforms trivially under the crystal symmetry\cite{annurev2015}. This is in sharp contrast with a multi-component CDW described by two order parameters $\psi_1$ and $\psi_2$ that transform differently under the crystal symmetries but share the same $\vec{q}_{CDW}$. Now, the collective Higgs mode ($\psi_1 \psi_2^* + \psi_1^* \psi_2$) will generally transform non-trivially under the symmetries of the crystal\cite{Fernandes2019}.

One example is the A$_2$ mode seen in the Raman response of \ch{URu2Si2}\cite{kung2015}, which was proposed to result from a previously hidden orbital ferroaxial state\cite{Hlinka2016}. Such states are rare and unlike other ferroic orders (Fig.~\ref{fig:symmetry-breaking}b). It is described as a circulating pattern of electric dipoles whose order parameter, an axial vector, preserves both spatial inversion and time reversal symmetries, and hence, does not couple linearly to electric and magnetic fields\cite{van2007,hayashida2020,dayroberts2024PRFA}. As such, the observation of an axial Higgs mode in \ch{GdTe3} and \ch{LaTe3}\cite{Nature606} was quite surprising as it indicated a hidden order whose specific type of ferroic density wave (Fig.~\ref{fig:symmetry-breaking}b) and thus origin have yet to be uncovered. Indeed, unlike other materials where unconventional density waves are believed to exist, the entire rare-earth tritellurides (\ch{RTe3}) have weak electron-electron correlations, no lattice frustration (e.g., as in Kagome systems), and in some cases, no $f$-electron or magnetic moments. Thus, the CDW in \ch{RTe3} had been considered conventional, possessing one or two unidirectional, incommensurate charge orders that only break some rotation and mirror symmetries related to the direction of $\vec{q}_{CDW}$\cite{fang2007stm, moore2010,chen2014,yumigeta2021,Yao2006,Johannes2008,Eiter2013} (Fig.~\ref{fig:symmetry-breaking}c). Interestingly, recent theoretical\cite{Alekseev2024} and experimental studies\cite{Que.CwSbTeSTM.2024} suggested the axial Higgs mode could emerge from a multi-component CDW involving orbital and charge order. In particular, the condensation of two CDW order parameters with different orbital compositions gave rise to a ferroaxial order\cite{Alekseev2024}. 

Beyond the type of unconventional CDWs, the universality of the axial Higgs across the \ch{RTe3} series is unclear. As shown in Fig.~\ref{fig:symmetry-breaking}f, rare-earth elements heavier than Gd can reduce T$_{CDW1}$ by half and produce a second subdominant CDW order below a lower temperature (T$_{CDW2}$)\cite{Ru2008,yumigeta2021}. Thus, for this study we turned to the heavier elements, which also provide a built-in control experiment wherein the broken symmetries of the system and Higgs modes emerging from different CDWs are measured in the same compound. Raman scattering reveals that the high-temperature order breaks all vertical and diagonal mirrors, producing an axial Higgs across the series. Consistent with a ferroaxial order emerging from orbital texture, electron diffraction confirms that the lattice mirror breaking is extremely weak. At the same time, the electron diffraction and rotational-anisotropy second-harmonic generation (RA-SHG) show the two-fold ($C_{2z}$)-rotation and inversion symmetries are preserved. Muon spin relaxation demonstrates the CDW state does not break time reversal. Within the same experiments on the same sample, we find that the low-temperature CDW only breaks an additional translational symmetry with respect to the high-temperature CDW phase, producing a scalar Higgs mode. Thus, the experimentally observed broken symmetries (Fig.~\ref{fig:symmetry-breaking}d) prove the axial Higgs mode results from an unconventional ferroaxial CDW. In addition, they provide compelling evidence of an electronic ferroaxial CDW emerging from a combined orbital and charge order. 

\section*{Amplitude modes of \ch{RTe3}}

\ch{RTe3} features two tetragonal Te-RTe-Te slabs separated by van-der-Waals gaps along the $c$-axis, Fig.~\ref{fig:symmetry-breaking}e, enabling exfoliation\cite{lei2020high,APL2019}. The $p$-orbitals of the Te square-net make up the Fermi surface with $p_{x}$ ($p_{y}$) parallel to the $a'$, 45$^\circ$ to {\it ${a}$}-axis ($b'$, 45$^\circ$ to {\it ${b}$}-axis) axis. The glide-plane along $b$-axis, between the Te-sheets (Fig.~\ref{fig:symmetry-breaking}e), produces slightly non-equivalent in-plane lattice parameters, resulting in a formally orthorhombic space group\cite{Ru2008,yumigeta2021}, but is weak enough that the normal state is effectively tetragonal\cite{singh2024,kivelson2023}. The CDW order has the wave-vector $\vec{q}_{CDW}=(2/7)b^*$ whose direction is selected by this effective glide field (Fig.~\ref{fig:symmetry-breaking}c)\cite{Ru2008,yumigeta2021, Straquadine2022}. Within the conventional scenario, as shown in Fig.~\ref{fig:symmetry-breaking}c, the unidirectional CDW order breaks translation, $C_4$-rotation and mirrors $m_{a'}$ (pink, 45$^\circ$ to {\it m$_{a}$}) and $m_{b'}$ (blue, 45$^\circ$ to {\it m$_{b}$}). However, for the ferroaxial CDW scenario, all mirrors parallel to the ferroaxial axis are broken, namely the vertical mirrors $m_{a}$ (red) and $m_{b}$ (green), as well as the diagonal mirrors $m_{a'}$ and $m_{b'}$ (Fig.~\ref{fig:symmetry-breaking}d). This order is unconventional, breaking symmetries beyond those of $\vec{q}_{CDW}$. Consequently, the system goes directly from a quasi-tetragonal to a monoclinic phase. 

\begin{figure}[ht]
\centering
\includegraphics[width=1\textwidth]{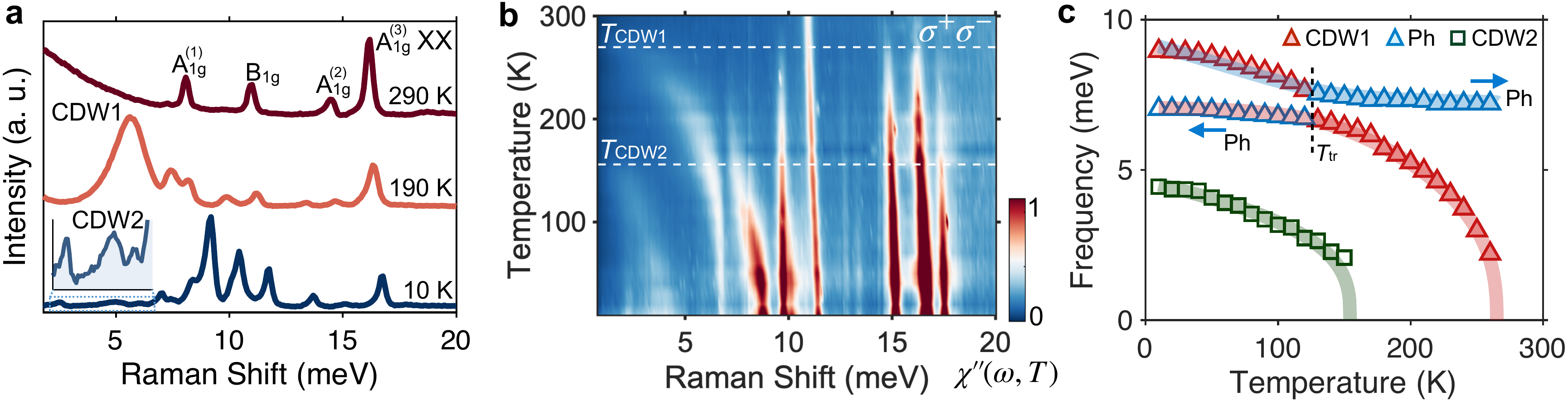}
\caption{ \textbf{Temperature-dependent Raman response of \ch{ErTe3}.} \textbf{a} Raman spectra of \ch{ErTe3} at 290 K, 190 K, and 10 K in parallel (XX) polarization configuration (the superscript in the mode symmetry at 290 K is used to distinguish three $A_{1g}$ modes). Inset shows the CDW2 AM (amplitude mode) and folded-phonons at 10 K. \textbf{b} Temperature-dependent Raman color map in the crossed-circular ($\sigma^{+}\sigma^{-}$) polarization configuration. The dashed white line indicates the CDW transition temperatures. \textbf{c} Ginzburg-Landau model fitting of the CDW2 amplitude mode as well as the coupled CDW1 amplitude mode and corresponding folded phonon mode. The solid red, blue, and green lines are fitted curves.}
\label{fig:temp}
\end{figure}

To simultaneously investigate the lattice and CDW symmetries, we carefully measured the polarization- and temperature-dependent Raman scattering in \ch{ErTe3}. Figure \ref{fig:temp}a shows the Raman spectra from the crystalline $a$-$b$ plane at 290 K (i.e., in the normal state \(T > T_{CDW1}\approx 265 K\)), 190 K (\(T_{CDW1} > T > T_{CDW2}\approx 155 K\)), and 10 K (\(T < T_{CDW2}\)) with the incident and scattered light polarized parallel to the $a$-axis (i.e., in the parallel (XX) configuration). In the normal phase (\(T > T_{CDW1}\)), four phonon modes are seen whose symmetries, discussed later, are consistent with a tetragonal, completely undistorted lattice. Upon entering the CDW1 phase, new modes appear either from the zone-folding by the CDW1 or due to the CDW Higgs mode\cite{APL2019, Nature606}. Upon further cooling below the second CDW transition temperature, additional weak modes at low energy are observed, suggesting further translation symmetry breaking (see the 10 K spectra in Fig.~\ref{fig:temp}a). 
 
The temperature evolution of most modes measured via the crossed-circular polarization Raman response ($\sigma^+$$\sigma^-$) is shown in Fig.~\ref{fig:temp}b. This and the other polarization channels (Supplementary Section 1) are consistent with previous measurements\cite{APL2019,maschek2018,Nature606}. As expected from zone-folded phonons, all but two modes appearing below \(T_{CDW1}\) are nearly temperature-independent. In contrast, the substantial energy shift of the 4.1 meV and 8.8 meV modes and their vanishing near \(T_{CDW2}\) and \(T_{CDW1}\), respectively, is consistent with the assignment of these excitations to the amplitude modes of the respective CDW orders\cite{PRB81,Eiter2013}. As shown in Fig.~\ref{fig:temp}c, the CDW2 amplitude mode (4.1 meV) is well described within a mean-field theory\cite{Gruner1994}, while the high-temperature CDW1 Higgs mode requires a two-mode interaction model\cite{PRB81,APL2019} to account for the avoided crossing at \(\sim\) 130 K with the phonon at \(\sim\) 7.5 meV (Supplementary Section 1). The coupling of CDW1 amplitude mode with the folded phonon (7.5 meV) reflects their identical symmetry. 

 \section*{Vertical and diagonal mirror symmetry breaking}

\begin{figure}[ht]
\centering
\includegraphics[width=1\linewidth]{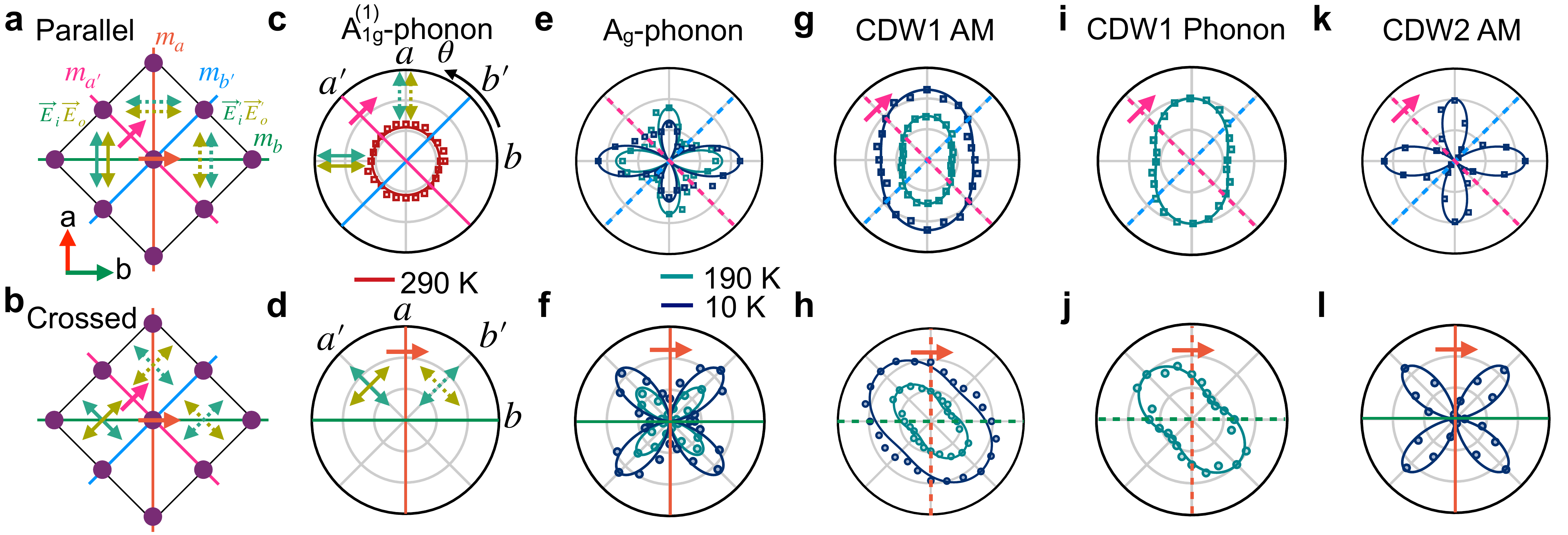}
\caption{ \textbf{Angular dependence of the phonon modes and CDW AM (amplitude mode) intensity in the parallel (XX)-(top) and crossed (XY)-(bottom) polarization configurations.} \textbf{a,b} Schematic illustrating the response of the mirrors in the parallel (a) and crossed (b) polarization configurations. Solid green and dark yellow arrows represent incident ($\vec{E}_{i}$) and outgoing ($\vec{E}_{o}$) polarization vectors, and dashed green and dark yellow arrows are the corresponding incident ($\vec{E}_{i}'$) and outgoing ($\vec{E}_{o}'$) polarization vectors under mirror transformation. \textbf{c,d} Angular response of the intensity of $A^{(1)}_{1g}$ (8.1 meV) phonon modes in XX \textbf{(c)}, and XY (not observed) \textbf{(d)} at 290 K (undistorted phase).  \textbf{e,f} Angular response of the intensity of $A_g$ (11.3 meV) phonon modes in XX \textbf{(e)}, and XY \textbf{(f)}, in the two CDW states at 190 K and 10 K (distorted phase). The response is consistent with the transition to an orthorhombic state with only breaking \textit{m$_a'$} and \textit{m$_b'$} mirrors. \textbf{g,h} CDW1 AM in XX \textbf{(g)} and XY \textbf{(h)}, at 190 K and 10 K, \textbf{i,j} CDW1 phonon (7.5 meV) in XX \textbf{(i)} and XY \textbf{(j)}, at 190 K (display two-fold behavior consistent with the axial response, where all the vertical (\textit{m$_a$}, \textit{m$_b$}) and diagonal (\textit{m$_a'$}, \textit{m$_b'$}) mirrors are broken). \textbf{(k,l)} CDW2 AM in XX \textbf{(k)} and XY \textbf{(l)} at 10 K respecting \textit{m$_a$} and \textit{m$_b$} mirrors consistent with $A_{g}$ symmetry. Solid curves are fitted with angular-dependence mode intensity equations as described in Table 1 of the Supplementary Section 2.}
\label{fig:CP}
\label{fig:Angular-Raman}
\end{figure}

We turned to angle-resolved Raman measurements to determine the symmetry of the high- and low-temperature CDW orders and their Higgs modes. It is helpful first to consider the vertical and diagonal mirrors of the Te square-net, shown in Fig.~\ref{fig:Angular-Raman}a,b as red ({\it m$_{a}$}) and pink ({\it m$_{a'}$}) lines. As shown in Fig.~\ref{fig:symmetry-breaking}c, since the $\vec{q}_{CDW}$ is along $b$-axis, in a conventional scenario the CDW would break the mirrors {\it m$_{a'}$} and {\it m$_{b'}$}, but not those normal or parallel to it (i.e., {\it m$_{a}$} and {\it m$_{b}$}). However, in the proposed case of unconventional CDW we anticipate all vertical and diagonal mirrors to be broken. To see how this affects the Raman response, consider that in \ch{RTe3} at our excitation energy, we excite from primarily the Te $p_{z}$ orbitals such that an $a'$ ($b'$) polarization selects a transition to $p_{x}$ ($p_{y}$). Hence, in the parallel polarization configuration, the Raman transitions occur between the same orbitals and, as such, are sensitive to {\it m$_{a',b'}$} mirrors that leave the orbitals unchanged. However, since crossed polarization configuration requires Raman transitions between orbitals, these are more sensitive to the {\it m$_{a,b}$} mirrors. 

Thus, we first evaluate the parallel polarization configuration in Fig.~\ref{fig:Angular-Raman}a, with an initial incident ($\vec{E}_{i}$, solid green arrow) and outgoing ($\vec{E}_{o}$, solid dark yellow arrow) electric fields along the $a$-direction (i.e., $aa$ (XX 90$^\circ$) configuration). The mirror {\it m$_{a}$} takes the $aa$ configuration to itself (dashed vertical arrows $\vec{E}_{o}'$ and $\vec{E}_{i}'$), whereas mirror {\it m$_{a'}$} takes the $aa$ to $bb$ (YY 0$^\circ$) configuration, horizontal arrows, $\vec{E}_{o}'$ and $\vec{E}_{i}'$. The mirror behaves distinctly in the crossed polarization configuration, shown in Fig.~\ref{fig:Angular-Raman}b. For example, for $\vec{E}_{i}$ along the $a'$ and $\vec{E}_{o}$ along $b'$ (i.e., $a'b'$ configuration), {\it m$_{a}$} swaps the $a'b'$ to $b'a'$ configuration, while under mirror {\it m$_{a'}$} they remain unchanged. Within this scenario, for the perfect square-net, one would expect equal Raman intensity under the {\it m$_{a}$} and {\it m$_{a'}$} mirror operations (similarly for {\it m$_{b}$} (green) and {\it m$_{b'}$} (blue)). Noting that in the orthorhombic phase from the conventional CDW, the {\it m$_{a'}$} ({\it m$_{b'}$}) mirror is broken, we expect distinct Raman responses for the $aa$ and $bb$ configurations. However, since {\it m$_{a}$} ({\it m$_{b}$}) maps $aa$ ($bb$) onto itself, the angular response in parallel polarization can produce a two-fold angular response (or two inequivalent two-fold responses along $aa$ and $bb$). In the crossed polarization, a four-fold angular response is expected, as preservation of the {\it m$_{a}$} and {\it m$_{b}$} mirrors forces the $a'b'$ and $b'a'$ response to be identical. 

Armed with this intuition, we note a tetragonal ($D_{4h}$ point group) symmetry will produce phonon modes with $A_{1g}$ and $B_{1g}$ symmetry in our scattering geometry. Figure \ref{fig:Angular-Raman}c,d shows the angular dependence of the intensity of the 8.1 meV phonon modes in the undistorted phase at 290 K. We observed angle independent intensity of the 8.1 meV in the parallel polarization configuration, Fig.~\ref{fig:Angular-Raman}c, and no signal in the crossed-polarization (Fig.~\ref{fig:Angular-Raman}d). This mode and those at 14.5 meV and 16.2 meV are well fit by the response expected of $A_{1g}$ symmetry (solid line in Fig.~\ref{fig:Angular-Raman}c) and thus labeled $A^{(1,2,3)}_{1g}$ respectively (superscript distinguishes the modes by energy). We also observed a phonon mode with $B_{1g}$ symmetry at 11.0 meV (see Supplementary Section 2). This is consistent with previous works\cite{Eiter2013,Lazarevic2011} and the normal phase having a tetragonal, $D_{4h}$ symmetry with all vertical ({\it m$_{a}$}, {\it m$_{b}$}) and diagonal ({\it m$_{a'}$}, {\it m$_{b'}$}) mirrors, as shown by solid lines in Fig.~\ref{fig:Angular-Raman}a,b, along with in-plane horizontal mirror and $C_{4}$-rotation. In the CDW1 phase (190 K), the phonon mode at 11.3 meV reveals unequal response along $aa$ and $bb$ parallel polarization configuration (Fig.~\ref{fig:Angular-Raman}e), confirming broken mirrors {\it m$_{a'}$} and {\it m$_{b'}$} (shown with dashed lines). However, the four-fold response in cross-polarization (Fig.~\ref{fig:Angular-Raman}f) suggests {\it m$_{a}$} and {\it m$_{b}$} remain (solid lines). Indeed, the observed angular response of this and another $B_{g}$ mode are well fit with the respective Raman tensors of the $D_{2h}$ point group (see Supplementary Section 2). Notably, no change in the angular response of these phonon modes is observed below \(T < T_{CDW2}\) (see Fig.~\ref{fig:Angular-Raman}e,f, dark blue), reflecting no additional symmetry breaking (except translation) due to the second CDW order. 

Distinct from the response of all other modes, the CDW1 amplitude mode (Fig.~\ref{fig:Angular-Raman}g,h) and the folded phonon mode at 7.5 meV (Fig.~\ref{fig:Angular-Raman}i,j) reveal a two-fold angular response in the parallel and, surprisingly, the crossed configurations. A similar symmetry of the two modes is expected as they mix and reveal an avoided crossing. Nonetheless, their anomalous response is identical to the axial Higgs mode in \ch{LaTe3} and \ch{GdTe3} (well fit with the antisymmetric Raman tensor \(R_{CDW} = \left( \begin{smallmatrix} a & d \\ -d & b\end{smallmatrix} \right) \)). The two-fold angular response in parallel polarization reflects the unidirectional nature of the CDW order with the charge fluctuations along the $a$-axis (wave-vector along the $b$-axis)\cite{Eiter2013} and is consistent with the breaking of the {\it m$_{a'}$} and {\it m$_{b'}$} mirrors. However, more uncommon is the response in the crossed-polarization configuration (Fig.~\ref{fig:Angular-Raman}h), which is also two-fold along the ${a'}$-direction reflecting the broken {\it m$_{a}$} and {\it m$_{b}$} mirror symmetries; consistent with a monoclinic rather than orthorhombic state. Since the folded phonon has the same response, we conclude that monoclinic symmetry exists in the electronic and lattice sectors. 

Next, we explore the symmetry of low-temperature CDW2 amplitude mode, where we expect a trivial angular response, as all vertical and diagonal mirror symmetries, as well as the ${C_4}$ rotational symmetry, are already broken. As shown in Fig.~\ref{fig:Angular-Raman}k,l, we find CDW2 amplitude mode exhibits a four-fold angular response in both parallel and crossed-polarization configurations, similar to the symmetric $A_{g}$ phonon mode (Fig.~\ref{fig:Angular-Raman}e,f). Thus, the CDW2 amplitude mode possesses a scalar representation, reflecting that the low-temperature order only breaks the additional translation symmetry without further lowering the point group. Moreover, the simultaneous detection of the axial (two-fold) high-temperature CDW1 and the scalar low-temperature CDW2 amplitude modes within the system rule out resonant or other Raman artifacts as the origin of the two-fold anisotropic CDW1 amplitude mode response. These findings are further confirmed by additional measurements performed on \ch{HoTe3} as well as \ch{ErTe3} measured both at Boston College (BC) and Massachusetts Institute of Technology (MIT) (see Supplementary Section 2). 

\begin{figure}[ht]
\centering
\includegraphics[width=1\textwidth]{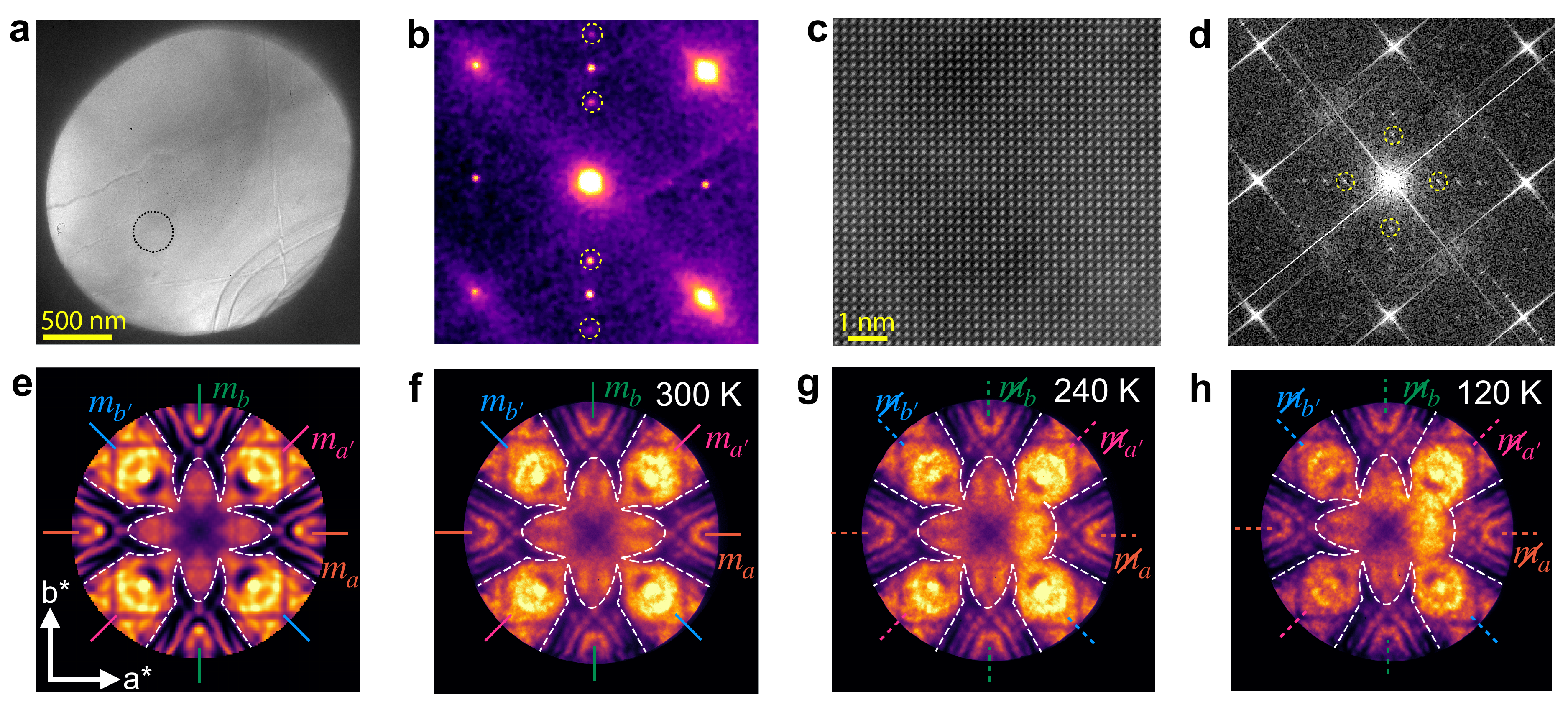}
\caption{\textbf{Large-angle convergent beam electron diffraction (LACBED) on \ch{ErTe3}.} \textbf{a} Bright-field TEM image of an \ch{ErTe3} flake. The black circle highlights the region from where the LACBED patterns were acquired. \textbf{b} SAED pattern taken on flake \textbf{(a)} at 240 K showing the satellite peaks (dotted yellow circles) from the high-temperature CDW1 phase. \textbf{c} Atomic-resolution HAADF-STEM image of \ch{ErTe3} taken at 120 K (bidirectional CDW state) in the [001] zone axis.\textbf{d} Fast Fourier transform of the atomic-resolution image \textbf{(c)}, dotted yellow circles indicate the satellite peaks from the two CDWs at 120 K. \textbf{e} Simulated LACBED pattern for a $\sim$ 46 nm thick \ch{ErTe3} flake without any broken symmetries, with the symmetric features highlighted by the overlaid dashed white lines. \textbf{f-h} Experimental LACBED patterns of \ch{ErTe3} flake at 300 K \textbf{(f)}, 240 K \textbf{(g)}, and  120 K \textbf{(h)}. Notice the changes in the dark features at the right half of the LACBED patterns below CDW1 transition temperature (indicated by dashed lines), reflecting the broken vertical ({\it m$_{a}$}, {\it m$_{b}$}) and diagonal ({\it m$_{a'}$}, {\it m$_{b'}$}) mirrors symmetries.}
\label{fig:STEM}
\end{figure}

To look for lattice changes more directly, we employed \textit{in-situ} cryogenic scanning transmission electron microscopy (STEM). As shown in Fig.~\ref{fig:STEM}b, the CDW1 transition was observed directly via the emergence of satellite peaks in the selected area electron diffraction (SAED). However, atomic-resolution STEM imaging using a high-angle annular dark-field detector (HAADF) at 120 K (Figs. \ref{fig:STEM}c,d) did not show the monoclinic distortion from the CDW transitions. This suggests that the monoclinic distortion seen with Raman scattering might be quite small. Thus, we turned to large-angle convergent beam electron diffraction (LACBED), which enhances the symmetry information encoded in conventional CBED patterns\cite{zuo2013} by extending the angular range of information in the reciprocal space by defocus (see Supplementary Section 4)\cite{tanaka1980,morniroli2002}. Since a crystal's structure governs how electrons are scattered, the symmetries of the crystal are reflected in the diffraction pattern\cite{buxton1976}, via variations in features such as lines, fringes, and intensity captured within LACBED. Figures \ref{fig:STEM}f-h show the central disk of LACBED patterns from an \ch{ErTe3} flake (shown in Fig.~\ref{fig:STEM}a) aligned along the [001] zone-axis at 300 K (\(T > T_{CDW1}\)), 240 K (\(T_{CDW1} > T > T_{CDW2}\)), and 120 K (\(T < T_{CDW2}\)), respectively. Key features are highlighted by white dashed lines, with the LACBED pattern at 300 K exhibiting a four-fold rotational symmetry. In addition, the image is symmetric when two sets of mirror planes perpendicular to the $<$100$>$ ({\it m$_{a}$}, {\it m$_{b}$}) and $<$110$>$ axes ({\it m$_{a'}$}, {\it m$_{b'}$}) are applied. These results are identical to the simulated LACBED pattern shown in Fig.~\ref{fig:STEM}e for tetragonal symmetry. Upon cooling below the CDW1 transition (Fig.~\ref{fig:STEM}g), in the right half of the LACBED pattern, the dark fringes smear out (overlaid dashed lines) while a noticeable enhancement in the intensity is observed. These changes in the LACBED pattern reflect the CDW1 order breaking of the mirror {\it m$_{a}$}, {\it m$_{b}$}, {\it m$_{a'}$}, and {\it m$_{b'}$} symmetries. No further symmetry change in the LACBED pattern is noticed with cooling below \(T_{CDW2}\), at 120 K (Fig.~\ref{fig:STEM}h). Supplementary Section 4 quantifies the symmetries in the LACBED patterns. 

\section*{Spatial inversion and time-reversal symmetries}

\begin{figure}[htp]
\centering
\includegraphics[width=0.5\linewidth]{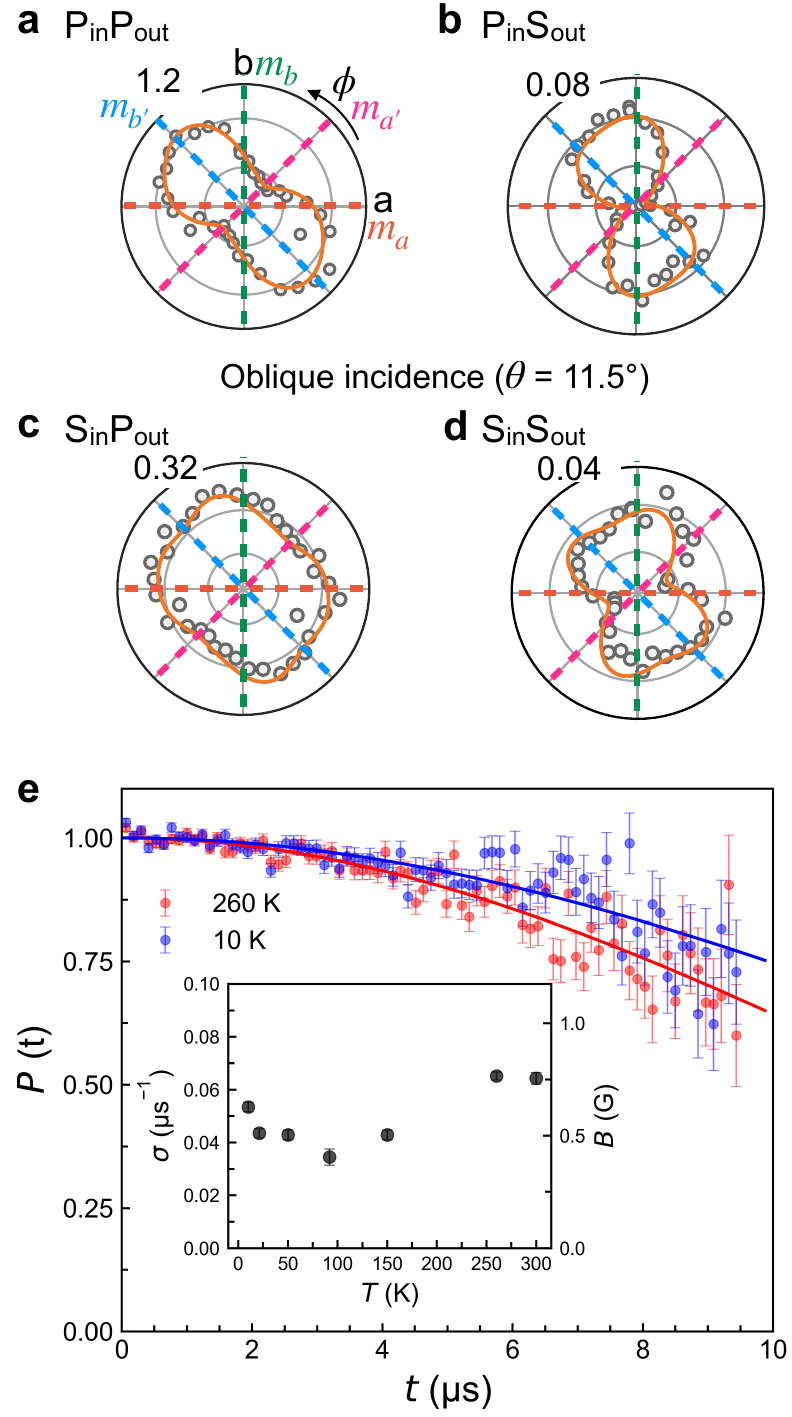}
     \caption{ \textbf{Broken symmetries in RA-SHG and $\mu$SR data of \ch{LaTe3}}. \textbf{a-d} Polar plots of RA-SHG data in four polarization channels, P$_{in}$P$_{out}$ \textbf{(a)}, P$_{in}$S$_{out}$ \textbf{(b)}, S$_{in}$P$_{out}$ \textbf{(c)}, and S$_{in}$S$_{out}$ \textbf{(d)} taken at 80 K on \ch{LaTe3}. Circles and solid lines denote the raw data and the fitting to the EQ RA-SHG under the $C_{2h}$ point group, respectively. The dashed lines in every plot indicate the original vertical and diagonal mirror planes. \textbf{e,} The ZF-$\mu$SR time spectra for \ch{LaTe3} obtained at 260 K (10 K) represented by solid red (blue) circles. The solid lines are the fittings of the data using the Gaussian Kubo-Toyabe function\cite{Kubo1967}. The inset shows the temperature dependence of the extracted relaxation rate, $\sigma$ and the equivalent width of the $B$-field on the right $y$-axis. The value is consistent with the La nuclear moment, reflecting no TRSB.}
     \label{fig:SHG-musr}
 \end{figure}

The experimental findings presented using Raman scattering and LACBED suggest the high-temperature CDW1 order is ferroaxial. However, these techniques do not probe the crucial spatial inversion and time-reversal symmetries needed to exclude other ferroic order parameters (e.g., polar or magnetic). Thus, we turned to RA-SHG, which detects broken inversion through electric dipole contributions and broken mirror symmetries in centrosymmetric phases through the electric quadrupole (EQ)\cite{Jin2020,Ahn2024}. Consistent with the absence of inversion symmetry breaking and the presence of out-of-plane $C_{2z}$-rotational symmetry, no SHG signal was observed in bulk \ch{ErTe3} and \ch{LaTe3} crystals near normal incidence. In \ch{ErTe3}, the RA-SHG was too weak at any temperature to clearly distinguish the symmetry breaking due to the CDW. Thus, we focused on RA-SHG in \ch{LaTe3} where a clear signal was resolved at an 11.5 angle of incidence from 80 K to 290 K, well below the CDW ordering temperature (\( T_{\text{CDW}} \approx\) 670 K). The RA-SHG polar plots in the four polarization channels at 80 K are shown in Fig.~\ref{fig:SHG-musr}a-d. Consistent with the Raman data, the broken mirror symmetries are directly reflected in the polarization dependent SHG response. Specifically, in the $P_{in}P_{out}$ configuration (Fig.~\ref{fig:SHG-musr}a), the response is not symmetric about the {\it m$_{a}$} and {\it m$_{b}$} mirrors, showing they are broken. More crucially, the responses of the three other polarization configurations, $P_{in}S_{out}$, $S_{in}P_{out}$, $S_{in}S_{out}$, do not respect any of the vertical and diagonal mirrors, showing the system is indeed monoclinic. Nonetheless, all four channels reveal a two-fold rotational symmetry, showing the highest possible symmetry point group for the CDW state is $C_{2h}$. Indeed, the EQ contribution to the SHG under $C_{2h}$ fits well with the experimental data, as shown by the orange lines in Fig.~\ref{fig:SHG-musr}a-d (Supplementary Section 5). Surveying fourteen locations across the sample and at other temperatures reveals results consistent with those in Fig.~\ref{fig:SHG-musr}a-d. Hence, the RA-SHG indicates the CDW state is centrosymmetric, requiring ferroaxial CDW order (Fig.~\ref{fig:symmetry-breaking}b).
 
To determine if the order breaks time-reversal symmetry, we employed muon spin relaxation ($\mu$SR), a sensitive probe of local magnetic fields\cite{Yaouanc2011}. Here, spin-polarized positive muons are implanted in the sample, and the time evolution of the $\mu$-spin polarization in the local magnetic field is measured via the preferential emission of a positron in the $\mu$-spin direction upon decay. The $\mu$SR from \ch{ErTe3} showed large fluctuating ${\text{Er}}^{3+}$ moment above 200 K, which do not exhibit long-range magnetic order, masking potential small contributions from orbital moments. Thus, we focus on the zero-field (ZF) $\mu$SR results from \ch{LaTe3} below \( T_{\text{CDW}}\), shown in Fig.~\ref{fig:SHG-musr}e. The lack of observable oscillations combined with the weak depolarization indicates the absence of long-range magnetic order. We fit the time-dependent curves with the Gaussian Kubo-Toyabe (GKT) function, which describes depolarization via densely packed, randomly oriented moments\cite{Kubo1967} (Fig.~\ref{fig:SHG-musr}e, see Supplementary Section 6). We can unambiguously rule out an increase in relaxation rate ($\sigma$) at low temperatures, the typical signature of time-reversal symmetry breaking (TRSB). If anything, there could be a slight decrease in $\sigma$ at low temperatures, which one may speculate is related to a change in muon-stopping site populations. Our stopping site calculations, see Supplementary Section 6, suggest that the observed relaxation is consistent with depolarisation due to the nuclear moments of $\ ^{139}\text{La}$. Based on these results, any ordering of novel electronic moments would produce a local field that must be less than the nuclear magnetic fields and is unlikely to be a source of TRSB. Finally, as discussed in Supplementary Section 3, the absence of TRSB is further supported by our helicity-resolved Raman response and the equivalent Stokes (S) and anti-Stokes (aS) Raman susceptibilities\cite{zhang2020, jin2020tunable,Huang2020,cenker2021,loudon1978,Yiping2020}.

\section*{Discussion}

Our extensive results demonstrate that the unconventional CDW1 in the entire \ch{RTe3} family breaks all vertical and diagonal mirrors but not inversion, C$_{2z}$-rotation, or time-reversal symmetries. Thus, the CDW involves a spontaneous breaking of multiple, distinct symmetries. Specifically, the CDW includes an out-of-plane ferroaxial moment that transforms as $A_{2g}$ in $D_{4h}$ (Fig.~\ref{fig:symmetry-breaking}a), lowering the system to $C_{4h}$. However, this combines with the uniaxial character of the CDW modulation, which transforms as $B_{g}$ in $C_{4h}$, further lowering the point group to $C_{2h}$. Thus, the axial Higgs mode results from an unconventional ferroic CDW transition, requiring the multi-component condensation of the CDW order parameters transforming as two different irreducible representations. Indeed, a conventional, single-component CDW could not lead to the emergence of a ferroaxial moment as it only breaks the symmetries associated with the single $\vec{q}_{CDW}$. This assertion is further corroborated by the amplitude mode of the low-temperature CDW having a scalar representation ($A_{g}$). Indeed, the second low-temperature CDW is conventional, only involving the further breaking of translation symmetry of the high-temperature phase. Furthermore, the  Raman scattering only reveals one phonon that directly manifests the monoclinic distortion. The modeling of the angular Raman response of the distinct single- and multi-domain ferroaxial state is discussed in Supplementary section 2. Thus, the Raman and STEM measurements reveal the CDW produces a nearly orthorhombic lattice distortion. However, the clear axial property of the amplitude mode and monoclinic space group seen in SHG suggests that the ferroaxial order has a much stronger imprint on the electronic sector. These results strongly suggest the ferroaxial component of the CDW results from an additional electronic component.

These findings align with a recent theoretical description emphasizing the role of both orbital and charge order in the \ch{RTe3} CDW\cite{Alekseev2024} that also appears to be substantiated by an STM study of \ch{CeSbTe}\cite{Que.CwSbTeSTM.2024}. Specifically, four of us revisited the microscopic model outlined in Ref. \cite{Yao2006}, where it was assumed that since $\vec{q}_{CDW}$ is along the Te square-net diagonal, the modulation of the occupation of the $p_{x}$ and $p_{y}$ orbitals was in phase. However, the Ginzburg–Landau model can be generalized to allow for a different CDW order parameter on the weakly hybridized $p_{x}$ and $p_{y}$ orbitals. Namely, within this microscopic model, the charge order pattern consists of two coexisting order parameters with the same $\vec{q}_{CDW}$, but distinct in either phase or amplitude. For negligible electron-phonon coupling, the model predicts a polar component in the CDW state (i.e., breaks inversion symmetry). However, the secondary order can become ferroaxial once the electron-phonon coupling is accounted for. If this ferroaxial CDW is realized, the extent to which this orbital-driven order is reflected in the electronic structure and its implications for quantum geometry remain open questions. Nonetheless, our experimental observations establish that a previously hidden multi-component unconventional ferroaxial CDW order and associated axial Higgs mode exists across the \ch{RTe3} series. As such, these systems may constitute a new class of density wave systems where unconventional order emerges purely from a strong coupling of the lattice, charge, and orbital degrees of freedom. These results further suggest measuring the properties of the Higgs modes in other CDW systems, such as dichalcogenides (\ch{TiSe2}) and kagome metals (\ch{CsV3Sb5}), would provide important insights into their novel ground states.

\newpage
\section*
{References}
\bibliography{References}

\newpage
\section*{Methods}
\subsection{Crystal growth}
High-quality ErTe$_3$ single crystals were grown in an excess of tellurium (Te) via a self-flux technique. Te (metal basis $>$ 99.999\% Sigma-Aldrich) was first purified to remove oxygen contamination and then mixed with rare earth ($>$ 99.9\%, Sigma-Aldrich) in a ratio of 97:3. The mixture was sealed in an evacuated quartz ampoule and heated to 900 °C over a period of 12 h and then slowly cooled down to 550 °C at a rate of 2 °C h$^{-1}$. The crystals were separated from the flux via centrifugation at 550 °C.

\subsection{Sample preparation and vacuum transfer}

Rare-earth tritellurides (\ch{RTe3}) are easily oxidized upon exposure. Thus it is crucial to keep these materials under an inert environment to prevent their oxidation. In our study, we have used freshly exfoliated samples for all the measurements. The samples were exfoliated onto \ch{SiO2}/Si substrates and characterized using unpolarized Raman spectroscopy within an argon-filled glovebox. Then, the exfoliated samples were transferred directly to a low-temperature cryostat under high vacuum (1.0 $\times$ $10^{-7}$ mbar) using a transfer suitcase\cite{ReviewofScientificInstruments}. As described previously,\cite{lei2020high,Nature606} this ensures the crystal orientation, atomically flat surfaces free from contamination and oxidation, as confirmed by TEM, and in Raman via the absence of Te oxide peaks and the polarization dependence.

\subsection{Raman Scattering}
The Raman measurements were performed with a custom-built low-temperature Raman set up in the backscattering configuration. The sample is excited with a 532 nm laser, and the laser power was kept low $(\leq 0.1~mW)$ to avoid local heating. The spectra were collected using an Andor spectrometer equipped with 2400 grooves per mm grating coupled with an Andor CCD detector, providing a resolution of approximately $1~cm^{-1}$. Temperature-dependent measurements were performed with a continuous closed-cycle He-flow cryostat (Montana Instruments). Angle-resolved Raman measurements at MIT were performed using Horiba LabRAM HR Evolution, while those at BC employed a custom-built low-temperature Raman setup\cite{tian2016low}. The incident and scattered light polarization were controlled using half-wave plates with a fixed analyzer while keeping the sample mounted inside the vacuum chamber. The spectra were averaged from three sets of measurements carried out in the same environment to ensure reproducibility. All Raman measurements were carried out on freshly exfoliated flakes.

\subsection{\textit{In-situ} cryogenic STEM measurements}

Single crystal \ch{ErTe3} was exfoliated onto a \ch{SiO2/Si} substrate using the Scotch® tape method in an argon-filled glovebox. \ch{ErTe3} flakes of desired thickness (\(\sim\) 46 nm) were identified using an optical microscope and then transferred to a MEMS-based TEM chip from DENS Solutions using polypropylene carbonate (PPC) polymer. The TEM chip was sealed inside the glove box and then transferred into a portable nitrogen-filled glovebag, inside which the TEM chip was loaded onto a HennyZ cryo-TEM holder capable of continuous temperature control from \(\sim\) 100-1000 K\cite{goodge2020}. The holder was loaded in an FEI Titan Themis S/TEM in the nitrogen environment. Thus, the \ch{ErTe3} flakes were never exposed to ambient atmosphere. The LACBED patterns were acquired at 60 kV in the nanoprobe mode, with a convergence angle of 30 mrad. The sample stage was first set to its eucentric height, where a converged beam of electrons is focused on the sample. The stage is then raised by \(\sim\) 3 $\mu$m above the eucentric height, which results in a small circular area (\(\sim\) 200 nm) of the \ch{ErTe3} flake to be illuminated by the electron beam, and a spot diffraction pattern is formed in the image plane. Next, we insert a selected-area aperture to block all diffracted beams, allowing only the direct central beam to form a LACBED pattern in the diffraction plane. 

\subsection{RA-SHG measurements}

RA-SHG measurements were taken under an oblique incidence geometry with an incident angle of $\theta$ = 11.5$^\circ$, incident fundamental wavelength of 800 nm, and reflected SHG wavelength of 400 nm. The laser is operated at a repetition rate of 200 kHz, a pulse duration of 50 fs at the sample site, and a focused beam full-width half maximum (FWHM) of 30 $\mu$m on the sample. A typical incident power is 1.0 mW, corresponding to a fluence of 0.7mJ/cm$^2$. RA-SHG records the SHG intensity as a function of an azimuthal angle $\phi$ between the light scattering plane and the in-plane $a$-axis, in four polarization channels, P$_{\mathrm{in}}$P$_{\mathrm{out}}$, S$_{\mathrm{in}}$P$_{\mathrm{out}}$, P$_{\mathrm{in}}$S$_{\mathrm{out}}$, and S$_{\mathrm{in}}$S$_{\mathrm{out}}$, where P/S$_{\mathrm{in/out}}$ stands for the polarization of the incoming/outgoing (in/out) light being parallel/normal (P/S) to the light scattering plane. The incident and reflected polarizations are defined by a half-wave plate and a linear polarizer, respectively, whereas the reflected SHG wavelength is filtered through two short pass and one bandpass filter with a total optical density (OD) = 18 for the fundamental wavelength. Due to the air sensitivity of \ch{LaTe3}, the samples were mounted into a cryostat inside a 99.999$\%$ N$_2$-filled glovebox to minimize the air exposure. The RA-SHG measurements were performed under a pressure less than 3 $\times$ 10$^{-7}$ mbar and across a temperature range between 80 K and 270 K. 

\subsection{$\mu$SR measurements}
The $\mu$SR measurements were carried out at the $\pi$M3 beamline using the general purpose surface-muon instrument\cite{gps} at the Paul Scherrer Institute, Switzerland. The single-crystalline samples of \ch{LaTe3} were mounted in a $^4$He gas-flow cryostat. We utilized the spin-rotated mode, where the muon spin is tilted approximately 45$^\circ$ with respect to the beam momentum. The sample is surrounded by three detector pairs: Forward/Backward, Up/Down, and Left/Right. The data presented in this text was from Forward and Backward detectors located along the horizontal beamline. The data were analyzed using the MUSRFIT software package\cite{musrfit}.

\subsection{Acknowledgement}
This work was primarily supported by the Air Force Office of Scientific Research under Grants No. FA9550-20-1-0282 (B.S., K.M.K., R.S., Y.W., L.M.S., and K.S.B.), FA9550-20-1-0260 (S.A.A.G., and J.C.) and FA9550-21-1-0423 (R.M.F.). Some Raman experiments were enabled by equipment provided through AFOSR DURIP award FA9550-20-1-0246. Work by GM and MG was supported by the National Science Foundation, Award No. DMR-200334 and DMR-2310895. PS and VMP are grateful for the support of the US Department of Energy (DOE), Office of Science, Office of Basic Energy Sciences under award number DE-SC0018675. The cryogenic STEM and LACBED experiments by SS, SDF and JJC were supported by the US DOE, Basic Energy Sciences program (DE-SC0023905). This work made use of the electron microscopy facility of PARADIM, which is supported by the NSF under Cooperative Agreement No. DMR-2039380, and CCMR Shared Facilities which are supported through the NSF MRSEC program (DMR-1719875). The FEI Titan Themis 300 was acquired through NSF-MRI-1429155, with additional support from Cornell University, the Weill Institute, and the Kavli Institute at Cornell. WZ and LZ acknowledges the support by the National Science Foundation through the Materials Research Science and Engineering Center at the University of Michigan, Award No. DMR-2309029. XL acknowledges the support by the AFOSR YIP grant no. FA9550-21-1-006.
JC is partially supported by the Alfred P. Sloan Foundation through a Sloan Research Fellowship and the Flatiron Institute, a division of the Simons Foundation. The muon work is based in part on experiments performed at the Swiss Muon Source S$\mu$S, Paul Scherrer Institute, Villigen, Switzerland.

\subsection{Author contributions}
BS performed the Raman experiments and analyzed the data. GM and KMK helped with data collection. RS and LMS grew the crystals. VMP and YW assisted in the setup of the automated low-temperature microscope. MG helped with sample preparation for Raman measurement. CAO and RC helped in Raman measurement. SS, SDF and JJC performed \textit{in-situ} cryogenic STEM measurements, WZ, XL, and LZ performed SHG measurements, PS, BS, TJH, CB, HL,and MJG performed the muon spin relaxation measurements. EMK helped in muon spin relaxation data analysis. SA, SAAG, JC, and RMF developed the theory. BS wrote the manuscript with the help of KSB. KSB and LMS conceived and supervised the project. All authors contributed to the discussion of the manuscript.

\subsection{Competing interests} The authors declare no competing interests. 

\end{document}